# Predicting the Porosity Formed in Freeze Casting by Artificial Neural Network

*Yue Liu, Wei Zhai, Kaiyang Zeng*

Department of Mechanical Engineering, National University of Singapore, 9 Engineering Drive 1, Singapore 117576
E-mail: mpezwei@nus.edu.sg

**Abstract:** Freeze casting has been increasingly applied to process various porous materials. A linear relationship between the final porosity and the initial solid material fraction in the suspension was reported by other researchers. However, the relationship of the volume fraction between the porosity and the solid material shows high divergence among different materials or frozen solvents, as the nature of materials significantly affects the pores formed in freeze casting. Here, we proposed an artificial neural network (ANN) to evaluate the porosity in freeze casting process. After well training the ANN model on experimental data, a porosity value can be predicted from four inputs which describe the most influential process conditions. The error range, reliability and optimization of the model were also analyzed and discussed in this study. The results proved that this method effectively summarizes a general rule for diverse materials in one model, which is difficult to be realized by linear fitting. Finally, a user-friendly mini program based on a well-trained ANN model is also provided to predict the porosity level for customized freeze-casting experiments.

## 1. Introduction

Freeze casting is a versatile and effective colloidal processing technique to produce porous materials, which has experienced fast-growth ever since the 2000s.[1] The technique starts with preparing a stable colloidal suspension, freezing the solvent in the suspension at a controlled low temperature or cooling rate, sublimating the frozen solvent from solid to gas state under reduced pressure, and finally sintering and densifying the remaining solid materials into porous materials.[2] Given the fact that the pores are mainly controlled by the frozen solvent, almost any types of solid materials, including ceramics, metals and polymers, can be produced into porous materials via freeze casting. To date, water, camphene and tertiary-butyl alcohol (TBA) are the most reported solvents, which provide a wealth of pore morphologies and structures. Freeze-cast porous materials are promising for a variety of applications including thermal and acoustic insulators,[3] scaffolds,[4–6] advanced composites,[7] gas separation membranes,[8] water filters,[9] electronic sensors[10] and packaging,[11] catalysts,[12] batteries,[13] *etc.*

Porosity is one of the key factors determining the functional and structural properties, such as sound absorption coefficient,[14] thermal conductivity,[15] mechanical strength,[16] *etc.*, of a porous material on top of the solid material characteristics. Freeze casting has been acclaimed as an effective way to control the porosity of the porous materials since the pores generated are, in principle, the replica of the frozen solvent.[2] The linear relationship between the porosity and initial solid materials volume fractions of freeze-cast porous materials have been reported by several researchers based on their process set-ups, specifically.[17–19] It leads to the possibility of predicting the porosity of a freeze-cast porous material prior to the freeze casting process, hence saving a considerable amount of time and resource for repeating experiments and processes. However, freeze casting is also considered as a complex process, where various parameters can affect the porosity: formulation of the suspensions (nature of the solid materials, solvents, particle sizes, binders, dispersants, *etc.*), freezing temperature, cooling rate, sintering temperature and time, *etc.*[1,2] Therefore, the complex interdependent relationships between porosity and these parameters cannot be concluded in a single, isolated paper, as almost every article has its own set of parameters.[1,2] In order to promote better informed experimental design, Dunand *et al.*[1] analyzed the relationship between the porosity of freeze-cast porous materials and solid material volume fractions among various materials based on more than 800 research papers. Though a linear relationship was proposed, huge diffractions were presented, leading to the difficulty in being used as a reference for the design of new experiments.[1]

Artificial intelligence (AI) method has a stronger capacity to describe the complicated relationship between variables than traditional statistical methods and empirical methods. One of its advantages is that AI can process different types of data synchronously without barriers.[20,21] Categorical data, numerical data and even images can be integrated into a dataset for a machine learning project. Besides classification and object recognition, numerical regression is one of the objectives which can be realized by machine learning.[22] In the field of materials science, AI, especially machine learning, has been a powerful tool for digging hidden information or potential rules.[23–27] As mentioned above, the volume fraction of the generated pores in freeze casting cannot be accurately predicted from the experimental conditions by simple linear regression because of the different solid materials and solvents characteristics, but it could be a typical problem the regression neural network is able to solve.

Here, we established an artificial neural network (ANN) to build a projection from the volume fraction of solid materials in the suspension to the porosity of the final freeze-cast porous materials, involving the effects of materials and solvents characteristics. It is a mixed-data neural network which is fed by categorical data and numerical data. The names of materials and solvents are encoded into categorical data, combined with the numerical variable—the volume fraction of solid in the suspension, as the inputs. The corresponding porosity can be quickly predicted from the mixed-data inputs by the ANN model. Our model is trained and tested by the data from real experiments of freeze casting. The architecture and configurations of the network are presented here in detail. In addition, the reasonability and stability of the model was also investigated in this study.

## 2. Results and Discussion
### 2.1. Predicting porosity by ANN

Artificial neural network (ANN) is a machine learning method which can realize supervised learning.[28] Supervised learning needs a prepared training set with input parameters and associated outputs. In this study, four sets of inputs are selected, which are the type of solid materials, the name of solid materials (*e.g.* $Al_2O_3$), the name of solvents, and the volume fraction of solids in the suspension. We chose three types of materials which are ceramics, metals and polymers, and three types of solvents which are water, tert-butyl alcohol (TBA) and camphene as inputs. The associated output is the porosity of the freeze-cast porous materials. A fully connected neural network was designed for seeking the relationship between the inputs and the final porosity. The schematic of the proposed ANN model is shown in **Figure 1**.

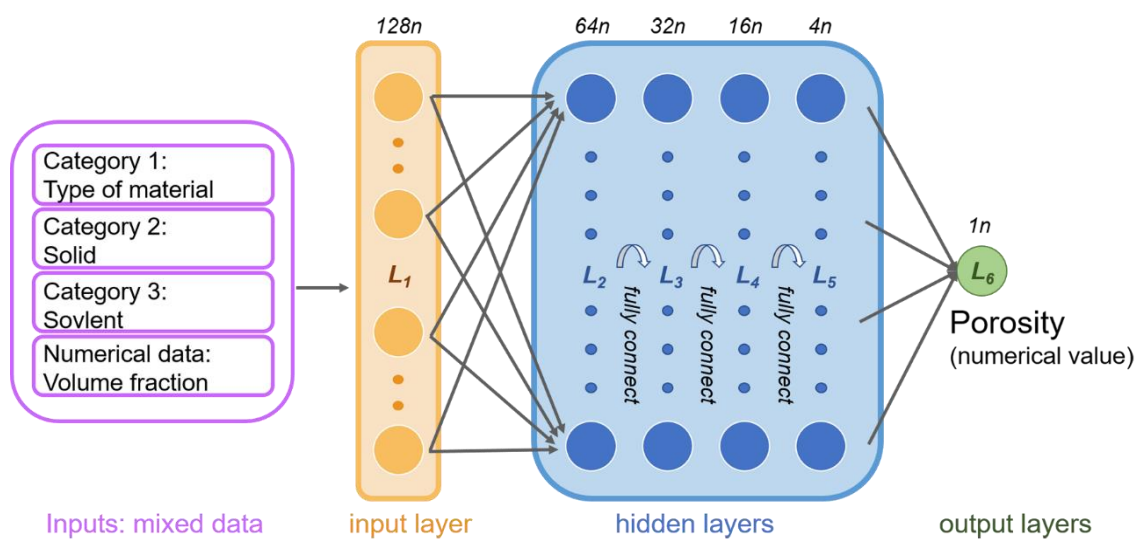

**Figure 1.** Structure of the artificial neural network (ANN) with three sections – one input layer, four hidden layers and one output layer. Four parameters from each freeze casting experiment are compacted into an array before feeding to ANN. The output is a numerical value representing the porosity level. All layers are fully connected, and the neurons in each layer are shown on the top (e.g. *128n*).

The input layer and output layer are arranged to adapt the dimension of the corresponding data. Each hidden layer consists of a number of artificial neurons. An activation function is attached to each neuron. The neurons in the first layer to the fifth layer use leaky rectified linear unit (Leaky ReLU) as their activation functions where the scalar multiplier for negative input values is set to 0.2. The linear activation function for regression is used in the last layer in order to get a numerical output. Each neuron is connected to all the neurons in the next layer. The connections have different weights. By calculating the weights in the hidden layers, a complicated mapping from inputs to output can be obtained.

**Table 1.** Comparison of the fitting parameters for the regression models describing the dependency of volume fraction on total porosity. $N$ is the number of samples, $r$ and $R^2$ are the coefficients of correlation and determination, respectively. For all the parameters in this table, $p < 0.0001$. *ns* indicates the value was insignificant.

|  | Reference[1] | | | Train + Test | | | Test | | |
| --- | --- | --- | --- | --- | --- | --- | --- | --- | --- |
|  | N | r | $R^2$ | N | r | $R^2$ | N | r | $R^2$ |
| ALL | 2855 | -0.12 | 0.02 | 2497 | **0.85** | **0.71** | 625 | **0.82** | **0.67** |
| SOLVENTS | | | | | | | | | |
| Water | 2044 | -0.66 | 0.43 | 1804 | 0.85 | 0.72 | 457 | 0.82 | 0.67 |
| Camphene | 290 | -0.61 | 0.37 | 275 | 0.81 | 0.65 | 71 | 0.8 | 0.64 |
| TBA | 408 | *ns* | *ns* | 418 | 0.86 | 0.72 | 97 | 0.81 | 0.65 |
| SOLIDS | | | | | | | | | |
| Ceramics | 2183 | -0.08 | 0.01 | 2168 | **0.81** | **0.65** | 532 | **0.77** | **0.59** |
| Metals | 131 | -0.72 | 0.51 | 127 | 0.77 | 0.58 | 38 | 0.75 | 0.51 |
| Polymers | 248 | -0.62 | 0.38 | 202 | 0.97 | 0.94 | 55 | 0.94 | 0.82 |

The prediction capacity of the ANN model should be validated on the test set and all data (training set and test set) in order to compare with Reference [1] (**Table 1**). If some small classifications are contained in the training set, it will cause the imbalance, negatively affecting ANN's learning capacity.[29] Hence, the solid materials with less than 5 samples and the solvents with less than 25 samples have been removed from the raw dataset. The Pearson correlation coefficient ($r$) is a measure of the linear correlation between two variables. The range of $r$ is [-1, 1], where -1 and 1 mean perfectly negative and positive linear correlation, respectively, and 0 is no linear correlation.[30] The coefficient of determination ($R^2$) is a measure describing how well the model predicts the dependent variable from the independent variable, ranging from [0, 1]. When $R^2$ is equal to one, it means 100% variation of the dependent variables can be described by the model.[31] For a certain material, the initial volume fraction of solid and porosity has a negatively linear correlation:[17,18,32]

$$\phi_p = a \cdot \phi_s + b, \tag{1}$$

where $\phi_p$ is the porosity, $\phi_s$ is the volume fraction of solid, $a$ is the slope, and $b$ is the intercept.

However, the regression fitting line cannot describe all the data well with a very small $r$ and $R^2$, as reported by Dunand et al.[1] When the regression line is obtained from the samples in the same material classification or using the same solvent, the fitting is improved. Less number of samples in the groups of metals and polymers is another reason for the improved linear fitting. By contrast, the ANN model performs well not only on ceramics, metals and polymers, but also on all data. The coefficients, $r$ and $R^2$, between the true value and the ANN-predicted value reach 0.85 and 0.71, much higher than -0.12 and 0.02 for the linear regression, respectively. It illustrates that, with the increasing data size, the general function **Equation 1** is failing, but ANN can consider the characteristics of materials and solvents to amend the relationship between the initial volume fraction of solid and the porosity of freeze-cast porous materials. It is a complicated regression which is difficult to realize by traditional statistical analysis, but ANN can learn it from the training data quickly.

The results on a test set are visualized and analyzed in detail to validate the ANN capacity. Analysis on each classification of materials is shown in **Figure 2**. There are 625 samples in this test set, including 532 ceramic samples, 38 metal samples and 55 polymer samples. The ANN-predicted porosity and true porosity have a high linear relationship, as $r$ and $R^2$ of them are 0.82 and 0.67, respectively [Figure 2(a)]. According to equation 1, the volume fraction of solid and the porosity in Figure 2(b) can be fitted to a straight line, $\phi_p = -0.847 \cdot \phi_s + 0.796$, by least squared method, where $r = -0.53$, and $R^2 = 0.28$. In total, we analyzed 76 materials, which include 54 of ceramics, 6 of metals and 16 of polymers. Different materials are marked by different colors in Figure 2(c), and the legends for Figure 2(c) are shown in Figure S1 (Supporting Information, SI). No specific material shows an obvious deviation in three classifications. The error distribution can be seen in Figure 2(d). Through the Kolmogorov-Smirnov (K-S) test,[33] the ANN-prediction error obeys the norm distribution for all data and each classification in the test set. *p*-values in the K-S test can be seen in Table S1 (SI). The norm distribution of the error also confirms the reasonability of our ANN model.

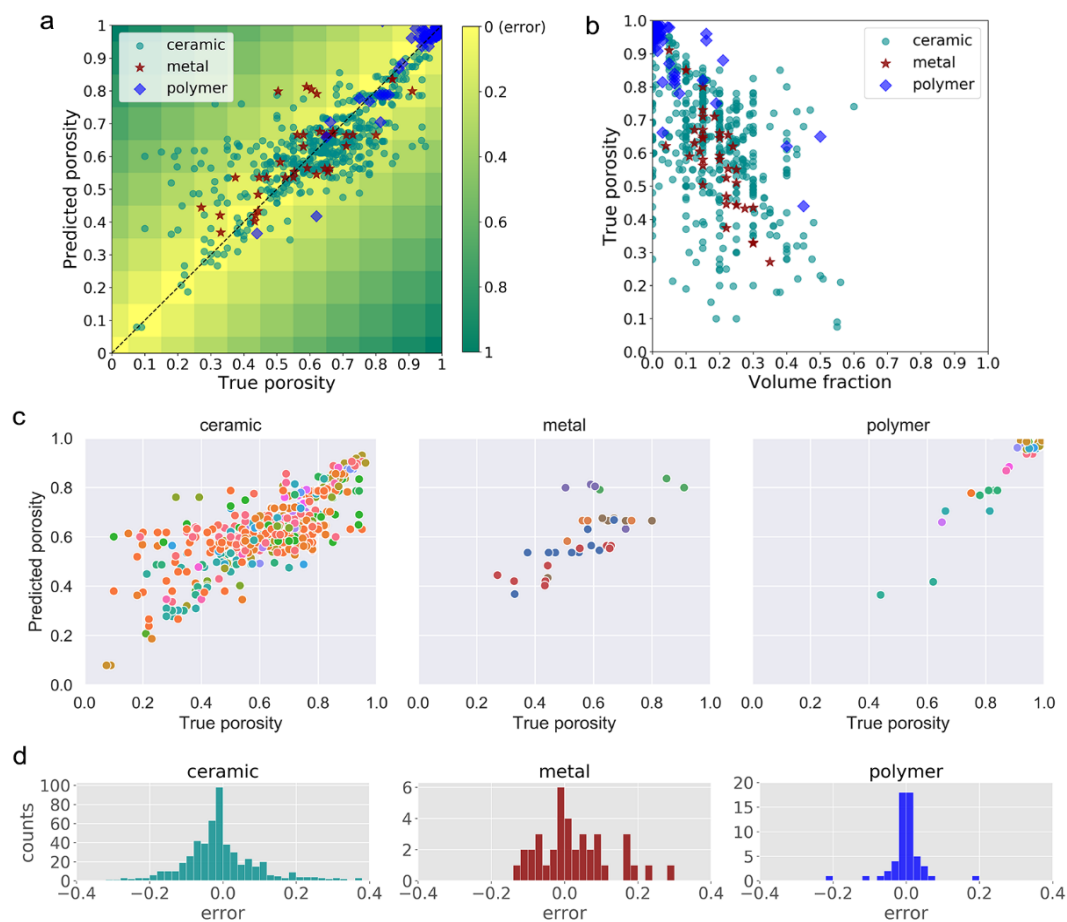

**Figure 2.** The performance for porosity prediction for different types of materials including ceramic, metal and polymer. (a) shows the comparison between ANN-predicted porosity and experimental value, and the background color illustrates the error value. (b) shows the relationship between volume fraction and true porosity. (c) and (d) illustrate the porosity prediction and errors for ceramic, metal and polymer, respectively. The legends of different colors in (c) are presented in Figure S2 due to a large amount of data.

The results can also be divided into three groups by different types of solvents (**Figure 3**). There are 457 samples in the group of water, 76 samples in the group of camphene, and 97 samples in the group of TBA. The errors distribution in three groups can be seen in Figure 3(c), which obtain $p \ll 0.0001$ in K-S test. Other evaluation parameters are shown in Table S1 (SI), which prove that the ANN-model has also well-considered the solvent characteristics in learning. In short, our ANN model shows a balanced and excellent performance for all the samples, no matter of which solid materials or solvents are involved.

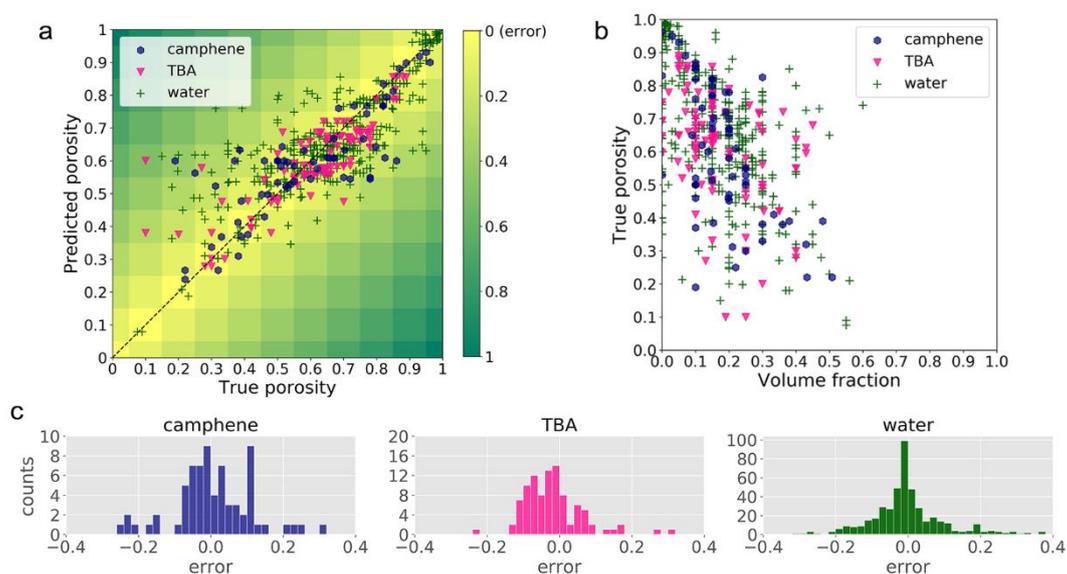

**Figure 3.** The ANN prediction performance for different solvents of camphene, TBA and water. (a) shows the comparison between predicted porosity and true porosity, and the background colors illustrate the error value. (b) shows the relationship between volume fraction and true porosity. (c) displays the histograms of prediction error.

### 2.2. Optimization of ANN

Designing the structure of ANN is flexible, and the optimal model is different for different tasks and data types. An unreasonable ANN model will limit the capacity of learning. Therefore, we compared and discussed the ANN models with different loss functions, widths and depths, and other configurations to determinate a reasonable architecture. In addition, the robustness is also confirmed by repetitive training.

*2.2.1. Loss function*

ANN aims at minimizing the error during the error propagation. Therefore, a loss function is necessary in ANN to calculate the error, or known as the loss, which can summarize the error of all training data down to a numerical value.[34] For a regression problem, there are four usual loss functions to choose from, mean squared error (MSE) (equation S1, SI), mean absolute error (MAE) (equation S2, SI), mean absolute percentage error (MAPE) (equation S3, SI) and root mean squared error (RMSE) (equation S4, SI). The evaluation of different loss functions can be seen in **Table 2**. We set the batch size as 12, the learning rate as $1e$-4 with a decay of $1e$-7, the neuron arrangement as Figure 1 showing, and the epoch as 1000 for all training in Table 2. The results indicate, except RMSE, there is little difference between MSE, MAE and RMSE. Comprehensively considering all fitting parameters, we chose MSE for our ANN model.

**Table 2.** Summary of the effects of loss functions on the ANN model's performance. $r$ and $R^2$ are the coefficients of correlation and determination, respectively. For all $r$ and $R^2$ in this table, $p < 0.0001$. MSE is mean squared error; MAE is mean absolute error; MAPE is mean absolute percentage error; RMSE is root mean squared error. MAE and MAPE are also regarded as the metrics of the fitting performance.

| Loss | $r$ | $R^2$ | MAE | MAPE (%) |
|------|-----|-------|------|----------|
| MSE  | 0.79 | 0.61 | 0.0764 | 15.83 |
| MAE  | 0.78 | 0.59 | 0.0762 | 15.97 |
| MAPE | 0.79 | 0.59 | 0.0775 | 15.07 |
| RMSE | Easy to get NaN loss during training | | | |

*2.2.2. Depth and width of ANN model*

The depth and width of an ANN model should be sufficient to process fine features of the data in order to obtain high accuracy. An intact ANN model has at least three layers which are an input layer, a hidden layer, and an output layer. With ANN becoming wider and deeper, the mapping from inputs to output can be more complicated to describe a more abstract relationship. However, an excessively deep ANN is hard to converge, also with problems like gradient explosion and overfitting.[34,35] In addition, the network will be redundant and slow, provided that the width of it is too large. Therefore, we compared the ANN models with different depths and widths in **Table 3**. We set the batch size as 12, the learning rate as 1$e$-4 with a decay of 1$e$-7, the loss function as MSE, and the epoch as 1000 for all training in Table 3. Note that MAE and MAPE can be minimized by a deep and wide enough ANN model (bolded in Table 3). While more neurons are arranged in more than six layers, the performance of the prediction cannot become better. After the systematic comparison, a six-layer ANN model with 128, 64, 32, 16, 4, 1 neurons in the first to last layer is chosen to process the data of freeze casting.

**Table 3.** Comparison of different amounts of layers and neurons on ANN performance. ANN structure is written as (neurons in Layer 1/…/neurons in Layer $n$). $r$ and $R^2$ are the coefficients of correlation and determination, respectively. For all $r$ and $R^2$ in this table, $p < 0.0001$. MAE is mean absolute error; MAPE is mean absolute percentage error.

| ANN structure | $r$ | $R^2$ | MAE | MAPE (%) |
|---|---|---|---|---|
| 32/16/1 | 0.79 | 0.62 | 0.0807 | 16.62 |
| 128/4/1 | 0.78 | 0.6 | 0.0808 | 17.04 |
| 128/32/1 | 0.79 | 0.62 | 0.0788 | 16.14 |
| 128/64/16/1 | 0.78 | 0.61 | 0.0784 | 16.6 |
| **128/64/32/16/4/1** | 0.79 | 0.61 | **0.0764** | **15.83** |
| 128/128/64/32/16/4/1 | 0.79 | 0.61 | 0.0779 | 15.79 |
| 256/128/64/32/16/4/1 | 0.79 | 0.62 | 0.0773 | 16.15 |

*2.2.3. Robustness*

Only a robust ANN model can describe an authentic and effective relationship between inputs and outputs.[36] Any eligible data can feed the network, and the inputs-output mapping should be similar. For the above tests (in section 2.2.1 and 2.2.2), we used a fixed state to split the training set and test set. After filtering, there are 2497 available samples from Reference [1]. It is reasonable that 75% of the samples are used for training the ANN model, and 25% are used to test the trained model. In order to confirm the robustness of our model, we randomly split the training set and test set by the ratio of 3:1 five times and compare the fitting parameters (**Table 4**). We set the batch size as 12, the learning rate as $1e$-4 with a decay of $1e$-7, the loss function as MSE, the neuron arrangement as Figure 1 showing, and the epoch as 1000 for all training in Table 4. The average $r$ is ~0.8, which is much higher than the coefficient between the volume fraction and the porosity ($r_{VF}$). The average MAE is ~0.07, which is stable in repetitive experiments.

**Table 4.** Summary of the robustness verification for the ANN model with different random state to split the train and test sets. $r$ and $R^2$ are the coefficients of correlation and determination, respectively. For all $r$ and $R^2$ in this table, $p < 0.0001$. MAE is mean absolute error; MAPE is mean absolute percentage error. $r_{VF}$ is the coefficient of correlation between the volume fraction and the porosity.

| Random State | ($N_{water}$/$N_{camphene}$/$N_{TBA}$) | ($N_{ceramic}$/$N_{metal}$/$N_{polymer}$) | $r$ | $R^2$ | MAE | MAPE (%) | $r_{VF}$ |
|---|---|---|---|---|---|---|---|
| 6 | (457/71/97) | (532/38/55) | 0.82 | 0.67 | 0.0725 | 15.78 | -0.52 |
| 18 | (475/58/92) | (538/34/53) | 0.83 | 0.69 | 0.0697 | 13.34 | -0.57 |
| 25 | (456/66/103) | (540/29/56) | 0.8 | 0.63 | 0.0781 | 15.49 | -0.54 |
| 34 | (449/60/116) | (537/30/58) | 0.78 | 0.59 | 0.0759 | 16.64 | -0.54 |
| 42 | (451/63/111) | (539/31/55) | 0.79 | 0.61 | 0.0764 | 15.83 | -0.56 |

*2.2.4. Other configurations*

In addition to the above experiments, other configurations in our ANN model were also investigated respectively. Learning rate is an important hyperparameter which affects the capacity and the speed of convergence. Attempts to set learning rate from 1$e$-7 to 1$e$-3 provide the best choice of 1$e$-4, which ensures the accuracy and meanwhile keeps a relatively high computing speed. We chose leaky ReLU[37] as the activation function for the first to fifth layer. Leaky ReLU with the scalar multiplier of 0.2 shows a slight advantage on prediction accuracy when comparing with ReLU and hyperbolic tangent (tanh). Adam is the most common optimizer for training ANN,[38] also benefitting our study. Batch size is a hyperparameter that defines the number of samples passed to the network for once updating the internal model parameters. It is essential to set a reasonable size for batches, or the learning capacity could be weakened. In this study, after various attempts, 12 samples in a batch make the ANN model learn better.

After 1000 epoch, most of the experiments we mentioned above show a good convergence, with stable losses in training set and test set. When the ANN model is designed and configured optimally, the loss decreases fast at first and then slowly tries to find a global minimum. The loss variation in 1000 epochs of the optimal ANN model is shown in Figure S2 (SI). The similar losses of the training set and the test set also indicate that the model does not overfit or underfit. Therefore, all models we mentioned in this study are trained for 1000 epochs.

## 2.3 Discussion

From the above evaluation, a good fit between the predicted data and the experimental data has been achieved. It shows that using inputs such as the type of solid materials, the name of solid materials, the name of solvents and the volume fraction of solids in the suspension, one can predict the porosity through the well-trained ANN model. A schematic diagram to indicate the key process steps and parameters to fabricate porous materials via freeze casting technique is shown in **Figure 4**. Ceramics, metals and polymers generally cover most types of materials that can be fabricated via freeze casting. Composites are normally produced through the infiltration of freeze-cast porous materials, resulting in dense materials, therefore, are eliminated from the current analysis. Different solid materials exhibit different chemical and physical properties, including thermal coefficient, volume change at different freeze and sintering temperature and rate, *etc.* In the freeze casting process, the generated pores are highly related to the solvent, as they are generally considered as the replica of the frozen solvent, as shown in Figure 4. Different types of solvent have different freezing points, volume change at different freezing temperatures, crystal structures, *etc.* Therefore, the materials classification, solid materials and solvents are key factors that can affect the final porosity of freeze-cast porous materials. The fourth input, volume fraction of the solid materials, refers to the volume of the solid materials against the total volume of the suspension. The total volume of the suspension includes the volume of the solid material, solvent, and the volume of other types of additives, such as binders, dispersants, anti-freeze additives, *etc.* During the freeze-casting process, nearly all the solvent and additives will be removed from the samples through freeze-drying and thermal process. The space taken by those frozen solvents and additives become pores in the final materials, as shown in Figure 4. All these contribute to the possibility and accuracy of predicting the final porosity of the samples using the volume fraction of solids in the suspension as input. As a result, we have included those four parameters as the inputs for the ANN model.

As discussed above, there are many other process parameters involved in the freezing casting experiments, such as the types of additives, freezing temperature, freezing rate, sintering temperature, time *etc.,* as shown in Figure 4. These factors also have a certain level of influence on the final porosity. For example, the volume change of solid materials at different sintering temperature and sintering time is different. Every article has its own setup of equipment and experimental environment. All these contribute to the diffraction in both linear fitting and ANN-prediction. For ANN-prediction, it limits the coefficients of correlation and determination to about 0.8 and 0.7, respectively, which are more acceptable than those of linear fitting. The most significant advantage of ANN is considering multiple variables with different data types in a model, which is hardly realized by statistical analysis. With reasonable configurations, the ANN model is robust. It proves that the general relationship, established by using ANN, between the final porosity of porous materials and the input parameters (the types of solid materials, the solid material name (*e.g.* $Al_2O_3$), the solvent name, and the volume fraction of solid in the suspension) is credible and stable. In other words, the well-trained ANN model can provide a reference value based on only four conditions after "a comprehensive consideration" which is learnt from massive experimental data.

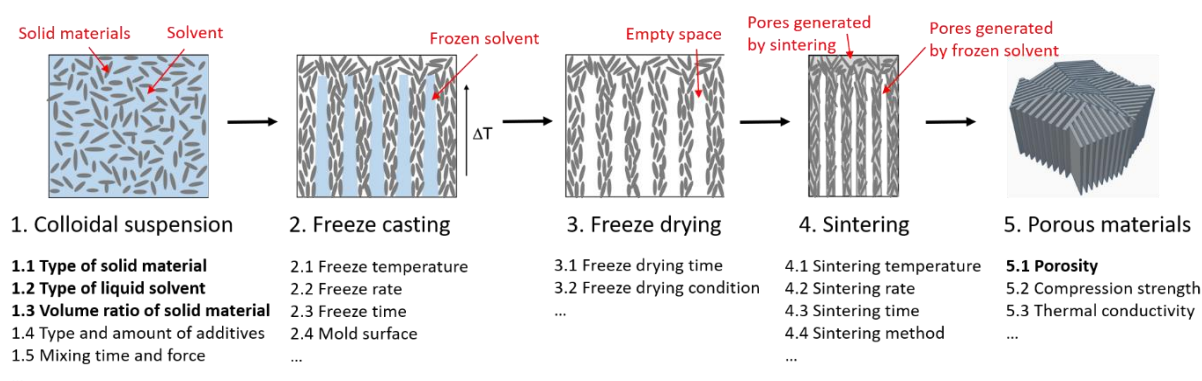

**Figure 4.** Schematic diagram of the key steps and process parameters to fabricate porous materials via freeze casting technique. There are four steps, including colloidal suspension preparation, freeze casting, freeze drying and sintering. For each step, there are certain key process parameters as indicated in the image.

Finally, we also designed a tool with a graphical user interface (Figure S3, SI). By inputting the information of the material on the panel, an ANN-predicted porosity can be obtained. It could help researchers optimize the experimental design and guide industrial applications, saving time and resources for the preliminary experiments.

**3. Conclusions**

In summary, we have developed an ANN method to predict the final porosity of freeze-cast porous materials, by choosing the solid and solvent involved and giving the volume fraction of solid material. Our ANN model is a fully connected neural network with six layers, requiring four inputs and giving one output. After learning from 1872 experimental samples, the ANN model captured the complex relationship between the inputting parameters and the porosity, and then the accuracy has been validated by a test set with 625 experimental samples. The dataset covered 76 materials including ceramics, metals and polymers, and the ANN model performed well on predicting all three types of materials. For the samples using different solvent of water, camphene, and TBA, there is no big diffraction of the prediction accuracy among them. The structure and configurations of the network have also been optimized by a series of experiments, and the robustness has been proved. On account of the reasonable error distribution and the systematic tests, we believe that the porosity value predicted by our ANN model has strong statistical significance and provides a credible reference for designing new experiments or industrial applications. Finally, a mini program with a graphical user interface has been attached here for easily setting a customized prediction. Therefore, it could be regarded as a general tool for predicting freeze-casting results, which is simpler and more intuitive than statistical analysis.

## 4. Method

**Data source.** The freeze casting experimental data was provided by an open-source database: FreezeCasting.net.[1,39]

**Materials involved.** After filtering the extremely small classifications, there were 2168 ceramic samples including 54 materials, 127 metal samples including 6 materials, and 202 polymer samples including 16 materials. 1806 of the samples used water as the frozen solvent, 275 of them used camphene, and 418 of them used TBA. The solid materials names refer to in Figure

S1 (SI). In total, there were 2497 samples, three quarters of them were randomly chosen to feed the ANN model for training, and one quarter of the total were used for testing the trained model.

**Statistical analysis.** In this study, all statistical methods applied to data have been described in detail in the main text, as well as the significant P values. All the statistical analysis was conducted by Python 3.6 with SciPy package.

**Code availability.** The ANN models were built in Python, mainly using open source libraries, Keras and Tensorflow. This ANN model does not take a long time to train. On a computer with Intel® Core™ i7-6500 CPU and 8GB RAM, the training process on the dataset with 1872 samples takes about 10 mins. Configurations of the model slightly influence the computing speed. The scripts are available from the corresponding author upon reasonable request.

## Supporting Information

The mini program can be download from: https://github.com/liuyuenus17/FreezeCasting-AI

Supporting Information.

## Acknowledgements

The authors acknowledge the support from National University of Singapore (NUS) with the Start-up grant "Microstructure-controllable template replication manufacturing of porous materials". YL would like to thank the research scholarship provided by NUS.

## Author Contributions

Y. Liu and W. Zhai conceived the method. Y. Liu performed the code work and statistical analysis. K. Y. Zeng provided supervision and technology advise. Y. Liu and W. Zhai wrote the manuscript.

**Conflict of Interest**

The authors declare no conflict of interest.

# Supporting Information

**Predicting the Porosity Formed in Freeze Casting by Artificial Neural Network**

*Yue Liu, Wei Zhai, Kaiyang Zeng*

Department of Mechanical Engineering, National University of Singapore, 9 Engineering Drive 1, Singapore 117576
E-mail: mpezwei@nus.edu.sg

**Table S1.** Fitting parameters corresponding to Figures 2 and 3. N is the number of samples in the group. $r$ and $R^2$ are the coefficients of correlation and determination, respectively. For all $r$ and $R^2$ in this table, $p < 0.0001$. MAE is mean absolute error; MAPE is mean absolute percentage error. KS $D$-value and $p$-value are obtained from Kolmogorov-Smirnov test.

| group name | N | r | $R^2$ | MAE | MAPE (%) | KS $D$-value | KS $p$-value |
|---|---|---|---|---|---|---|---|
| ALL | 625 | 0.82 | 0.67 | 0.0725 | 15.78 | 0.40 | 2.70E-91 |
| water | 457 | 0.82 | 0.67 | 0.0719 | 14.69 | 0.41 | 1.91E-68 |
| camphene | 71 | 0.80 | 0.64 | 0.0832 | 18.49 | 0.40 | 1.25E-10 |
| TBA | 97 | 0.81 | 0.65 | 0.0674 | 18.96 | 0.43 | 3.72E-17 |
| ceramic | 532 | 0.77 | 0.59 | 0.0770 | 17.15 | 0.40 | 1.48E-76 |
| metal | 38 | 0.75 | 0.51 | 0.0757 | 14.54 | 0.45 | 2.01E-07 |
| polymer | 55 | 0.94 | 0.82 | 0.0267 | 3.37 | 0.45 | 7.18E-11 |

**ceramics**

Material
- Bioglass
- Mullite
- YSZ
- Al2O3
- HAP
- SiC
- kaolin
- Al2O3-SiC (layered)
- Mullite-Na2SO4 (impregnation)
- YSZ-40wt.% LSM (mixed)
- TiO2
- Al2O3-ZrO2 (mixed)
- Al2O3-0.2wt.% ZrO2 (mixed)
- TCP
- HAP-SiO2 (mixed)
- Si3N4
- BCP-20wt% HAP (mixed)
- LSM-YSZ (layered)
- SiO2
- cordierite
- CaP
- BaTiO3
- ZrB2
- Al2O3-18wt.%ZrO2 (mixed)
- Al2O3-7wt.% ZrO2 (mixed)
- Merwinite
- PZT
- SiAlON
- Al2SiO5
- Li(NiCoAl)O2
- Y2SiO5
- BT-HAP (mixed)
- Al2O3-0.44.%ZrO2 (mixed)
- HAP-5wt.% SiO2 (mixed)
- BCP
- fireclay
- Al2O3-mullite (mixed)
- Mullite-2wt% ZrO2
- Al2O3-ZrO2 (layered)
- SnO2
- NiTiNbO4
- Zeo
- SiOC
- CaSiO3
- Ni0.5Ti0.5NbO4
- ZrO2
- ZrB2-12wt.%SiC-3wt.%Si3N4 (mixed)
- Cr3C2
- Silicate
- palygorskite
- Yb2SiO5
- glass

**metals**

Material
- Ti, 3.8wtO2
- Ti6Al4V
- Ni
- W
- Fe
- Cu

**polymers**

Material
- PU
- PHBV
- TEMPO-oxidized cellulose
- chitosan
- Na-MMT-Cellulose
- Enzymatic-cellulose
- PVA
- HEMA
- latex
- chitosan-PMMA (mixed)
- gelatin
- cellulose
- VGCF-PTMA
- silk fibroin
- PLLA

**Figure S1.** Legends of Figure 2c.

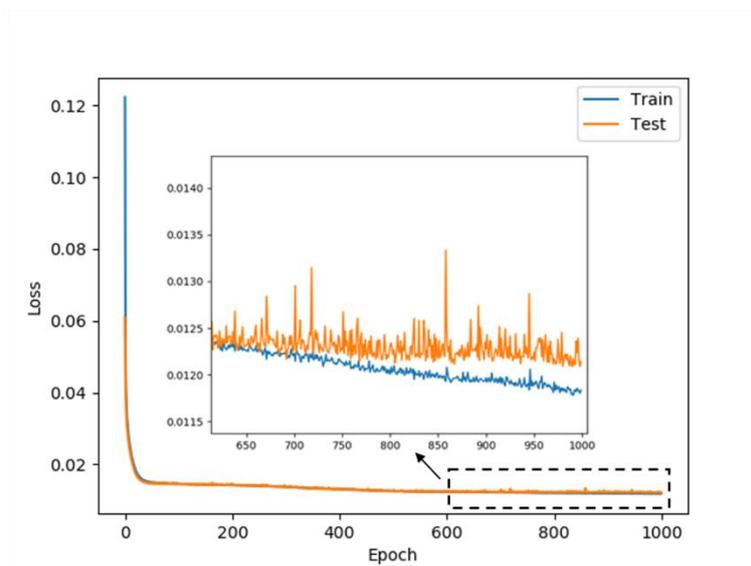

**Figure S2.** Loss decrease during ANN model training.

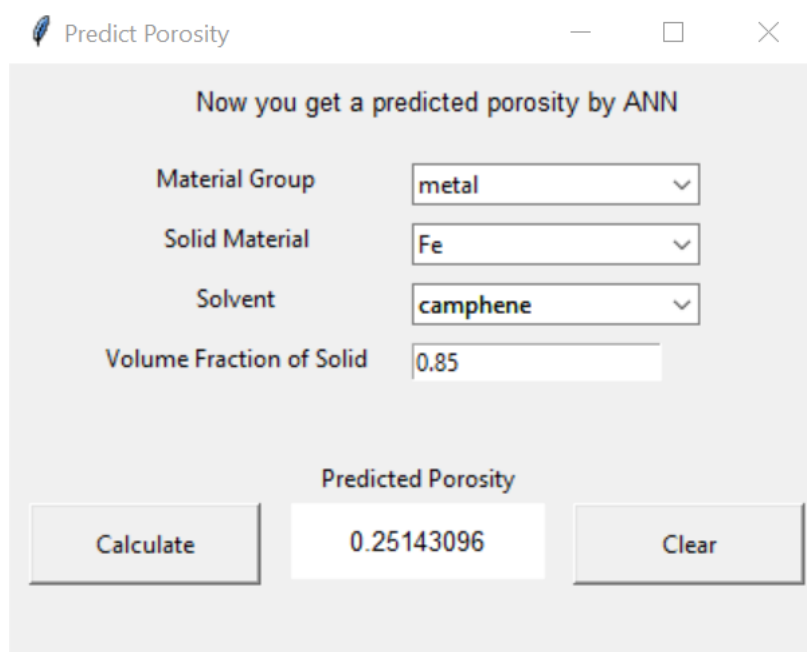

**Figure S3.** The interface of the ANN-prediction tool.

**Loss functions:**

$$\text{MSE} = \frac{1}{n}\sum_{j=1}^{n}(y_j - \hat{y}_j)^2, \tag{S1}$$

$$\text{MAE} = \frac{1}{n}\sum_{j=1}^{n}|y_j - \hat{y}_j|, \tag{S2}$$

$$\text{MAPE} = \frac{100\%}{n}\sum_{j=1}^{n}\left|\frac{y_j - \hat{y}_j}{y_j}\right|, \tag{S3}$$

$$\text{RMSE} = \sqrt{\frac{1}{n}\sum_{j=1}^{n}(y_j - \hat{y}_j)^2}, \tag{S4}$$

where $y$ is the observed value and $\hat{y}$ is the predicted value, $n$ is the total number of the samples in a set.